\documentclass[letter,traditabstract]{aa}

\usepackage{graphicx}
\usepackage{txfonts}
\def\mstar  {$M_{\star}$}
\def\macc   {$\dot{M}_{\rm acc}$}
\def\mdisk {$M_{\rm disk}$}

\def\nodata {...}

\begin{document} 

\title{Why do protoplanetary disks appear not massive enough\\ to form the known exoplanet population?}
\titlerunning{Planet and disk masses}
\authorrunning{Manara et al.}

   \author{C.F. Manara\fnmsep\thanks{ESO Fellow}
          \inst{1}
          \and
          A. Morbidelli\inst{2}
        \and
          T. Guillot\inst{2}
          }

   \institute{European Southern Observatory, Karl-Schwarzschild-Strasse 2, 85748 Garching bei M\"unchen, Germany\\
              \email{cmanara@eso.org}
         \and
             Laboratoire Lagrange, Universit\'e C\^ote d'Azur, Observatoire de la C\^ote d'Azur, CNRS, Boulevard de l'Observatoire, CS 34229, 06304 Nice Cedex 4, France
             }

   \date{Received August 13, 2018; accepted September 19, 2018}

 
  \abstract{When and how planets form in protoplanetary disks is still a topic of discussion. Exoplanet detection surveys and protoplanetary disk surveys are now providing results that are leading to new insights. We collect the masses of confirmed exoplanets and compare their dependence on stellar mass with the same dependence for protoplanetary disk masses measured in $\sim$1--3 Myr old star-forming regions. We recalculated the disk masses using the new estimates of their distances derived from Gaia DR2 parallaxes. 
We note that single and multiple exoplanetary systems form two different populations, probably pointing to a different formation mechanism for massive giant planets around very low-mass stars.
While expecting that the mass in exoplanetary systems is much lower than the measured disk masses,  we instead find that exoplanetary systems masses are comparable or higher than the most massive disks. This same result is  found  by converting the measured planet masses into heavy element content (core masses for the giant planets and full masses for the super-Earth systems) and by comparing this value with the disk dust masses. Unless disk dust masses are heavily underestimated, this is a big conundrum.
An extremely efficient recycling of dust particles in the disk cannot solve this conundrum. 
This implies that either the cores of planets have formed very rapidly ($<$0.1--1 Myr) and a large amount of gas is expelled on the same timescales from the disk, or that disks are continuously replenished by fresh planet-forming material from the environment. These hypotheses can be tested by measuring disk masses in even younger targets and by better understanding if and how the disks are replenished by their surroundings.
}


   \maketitle
%
\section{Introduction}
At least 30\% of stars have planets \citep[e.g.,][]{zhu18} and, given current detection limits, it is plausible that planetary systems exist around every star. Similarly, in clusters younger than 2 Myr, $\gtrsim$60--80\% of stars possess protoplanetary disks \citep[e.g.,][]{fedele10}. Determining when these planets formed should be simple. It would require comparing the median mass of exoplanetary systems to the decrease in protoplanetary disk mass with age and determining the intersect. However, the lack of large samples of
protoplanetary disk mass measurements and the incompleteness of exoplanet detection surveys have long hindered the task of constraining planet formation timescales. Furthermore, the fact that the millimeter emission from protoplanetary disks traces only the small dust mass (less than $\sim$cm)  and that total disk masses are still uncertain \citep[e.g.,][]{BW18} makes the comparison more challenging.

Works based on surveys of protoplanetary disk masses carried out before the advent of the Atacama Large Millimeter/submillimeter Array (ALMA) and on initial results from radial velocity and transit planet detection surveys, indeed suggested that the mass of protoplanetary disks is usually lower than the mass of the detected giant planets. In particular, \citet{GR10}, \citet{williams12}, and \citet{NK14} showed that only a tiny fraction of disks contain enough mass to explain the mass of the large number of observed gas giant planets in the standard core accretion model. They proposed that planet formation must have been underway by the time these disks were observed ($\sim$1--3 Myr). Similarly, \citet{mulders15} and \citet{mulders18} used the planet detections from the Kepler mission to show that the measured mass in disks around low-mass stars is typically smaller than the observed amount of heavy elements  in the planetary systems around the same kinds of stars. 

The time is now ripe to compare the masses of disks and planetary systems for a wide range of stellar types.  Surveys of disks in several star-forming regions carried out with ALMA over the past few years, combined with optical to near-infrared spectroscopic surveys, are showing that the mass of the dust content of protoplanetary disks increases with stellar mass following a steeper than linear relation and decreases with time \citep[e.g.,][]{ansdell16,pascucci16,barenfeld16,ansdell17}. These results also show that disks around  low-mass stars are less massive than the total mass in exoplanetary systems around similarly low-mass stars, for example TRAPPIST-1 \citep{pascucci16, testi16}.  At the same time, surveys of exoplanets have now detected planets around stars with different stellar masses, and the dependences of the planet properties on host star masses can be addressed \citep[e.g.,][]{mulders18}.

Here we perform for the first time a detailed comparison of the disk masses measured with ALMA in young star-forming regions with the current information on planetary systems masses, as well as core masses,    
to infer information on planet formation timescales and processes.

\section{Sample and data}

Exoplanet data are taken from exoplanet.eu \citep{schneider11}. From the catalog of 2018 July 10 we select only the confirmed planets. Since our analysis is based on the dependence of the stellar mass (\mstar) to the planet mass ($M_{\rm pl}$), we only select targets for which these values are both available. 

\begin{figure}[]
\centering
\includegraphics[width=0.45\textwidth]{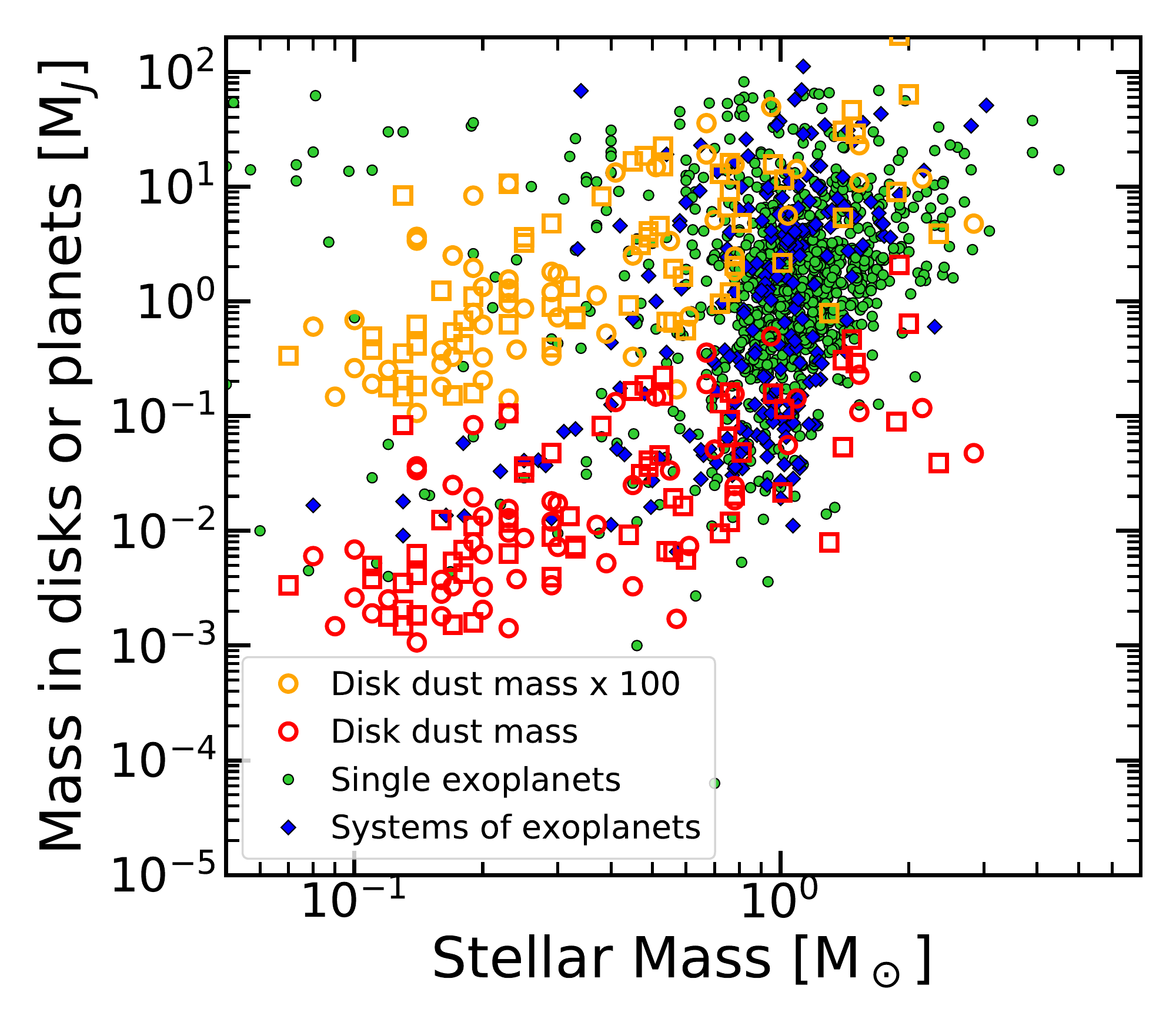}
\caption{Masses of single exoplanets, exoplanetary systems, and disk masses as a function of the mass of their host star. Empty squares are used for disks in the Chamaeleon~I region, while empty circles for disks in the Lupus region.
     \label{fig::mplan_mdust_mstar}}
\end{figure}

Starting from the catalog of confirmed exoplanets we construct our catalog of masses of exoplanetary systems by summing up the masses of individual planets in each system. These planetary masses are directly measured for $\sim$80\% of the exoplanetary systems. The reported masses represent a lower limit to the total mass in a system since planet detection surveys are still incomplete and a factor of 1/sin($i$) makes the actual masses larger most of the times.  
The masses of exoplanetary systems  discussed here are still expected to be modified by future exoplanet detections. On the one hand, the masses of exoplanetary systems  will increase when multiple planets in the known systems are detected. This will be a small or negligible effect, since only the lower mass planets are still unknown in a given exosystem. On the other hand, more sensitive exoplanet surveys will detect less massive planets, perhaps revealing the existence of other exoplanetary systems with overall lower total planet masses. These combined effects will increase the already wide (4 dex) spread of exoplanetary systems masses at any given stellar mass (see Fig.~\ref{fig::mplan_mdust_mstar}). However, the current data already allow us to draw  reasonable preliminary conclusions on when and how planets form, as discussed in the following. 

The measurements of protoplanetary disk masses (\mdisk)  are obtained from the surveys carried out with ALMA for disks in the $\sim$1--3 Myr old Lupus and Chamaeleon~I star-forming regions, the youngest regions extensively studied with ALMA to date \citep{ansdell16,pascucci16}. 
Disk dust masses are derived in these surveys by converting the millimeter continuum flux using a single value for the dust opacity and the disk temperature. Such measurements are thus only sensitive to grain sizes up to $\sim$cm and are based on the emission from the outer regions of disks ($R>10$ au) \citep[e.g.,][]{BW18}.
Disk total masses, inferred from gas emission lines, are still more uncertain \citep[e.g.,][]{miotello17,BW18}. For this reason, instead of using gas emission lines data, we estimate the total disk mass starting from the dust mass in disks and assuming a constant gas-to-dust ratio of 100.
For these regions \mstar \ is available for the vast majority of the disk-hosting stars from the spectroscopic surveys of \citet{alcala14,alcala17} in Lupus, and of \citet{manara16a,manara17a} in Chamaeleon~I. As discussed in Appendix~\ref{sect::dr2}, we rescale the disk masses and stellar luminosity, and thus recalculate stellar masses, using the distances inferred from the newly delivered Gaia DR2 parallaxes \citep{gaia,gaiadr2}. The availability of accurately determined stellar masses is needed for our analysis. For this reason we cannot include the results of the survey in the $\rho$-Ophiucus region by \citet{cox17}.

The dependence of both the exoplanetary systems, single exoplanets, and disk masses on the host star mass is shown in Fig.~\ref{fig::mplan_mdust_mstar}.  
As noted by  \citet{pascucci16}, among others,  the dependence of disk masses on stellar masses for the targets in the Lupus and Chamaeleon~I regions has a steeper than linear slope and the two distributions are indistinguishable, although the disks in the Lupus region are slightly more massive than those in the Chamaeleon~I region \citep{ansdell17}. In the following, we consider the data from the two regions altogether. 
The main difference between the distribution of the masses of single exoplanet and systems of exoplanets is the population of very massive single exoplanets ($M_{\rm pl}\gtrsim M_J$)  around stars with $M_\star<0.3 M_\odot$, a region of the parameter space where no exoplanetary systems (i.e., only single planets) have been found to date. 
Since the mass ratio of these massive planets to their (small) host star is high ($10^{-2}$ to $\sim 1$), it is plausible to infer that they may have been formed directly, as in binary stellar  systems.
In the following, we consider mainly the total masses in multiple exoplanetary systems for the comparison with the disk masses.

\section{Comparison between disk and exosystem masses}\label{sect::results}
\begin{figure}[]
\centering
\includegraphics[width=0.45\textwidth]{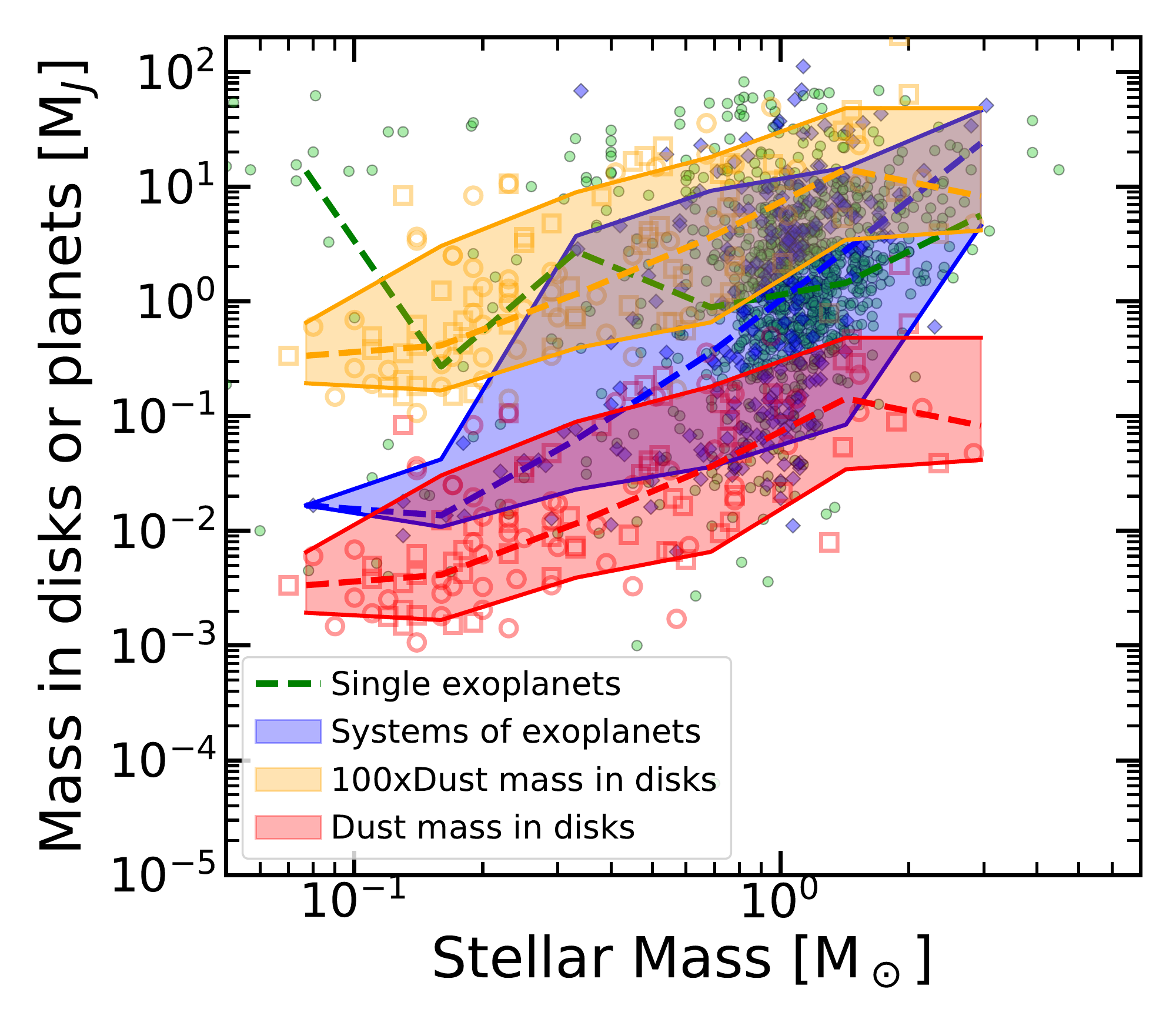}
\caption{Masses of single exoplanets, exoplanetary systems, and disk masses as a function of the mass of their host star, as in Fig.~\ref{fig::mplan_mdust_mstar}. The colored regions encompass the 10th and 90th percentiles of the distributions, while the dashed lines represent the median of the distributions, as labeled.
     \label{fig::mplan_mdust_mstar_perc}}
\end{figure}

When comparing the total mass in exosystems with the disk masses, we note that for low stellar masses ($M_\star\lesssim0.3 M_\odot$) the planetary system masses increase with  stellar mass as the disk dust  mass, although with a weak correlation. The median of the disk dust masses in this stellar mass range is a factor of $\sim$0.2--0.3 of the exoplanetary system masses (Fig.~\ref{fig::mplan_mdust_mstar_perc}), smaller than the expected $>$1 factor. It is possible that smaller planets are more difficult to detect, but since the upper envelope (90th percentile) of the distribution of disk dust masses is barely compatible with the median of the exosystem masses, the existence of smaller mass planets would not change the fact that there are planets more massive than the most massive disks.  This means that exoplanetary system masses are usually higher, and in the best-case scenario comparable with disk dust masses. In this stellar mass regime planets have $M_{\rm pl}\lesssim 10 M_\oplus$, and they can be considered to be mainly rocky; instead, for higher stellar and planetary system masses, a wide spread of planet masses is present, with low-mass rocky planets as well as higher mass planets with gaseous envelopes detected. When comparing the planet masses with the total disk masses under the assumption of a constant gas-to-dust ratio of 100, it is found that the median of the disk masses is higher than the median of the exoplanet masses by a factor of less than 10, while the 90th percentile of the two distributions are always comparable within a factor of 3. 
However, the assumption of a constant gas-to-dust ratio of 100 is uncertain. 

\begin{figure}[]
\centering
\includegraphics[width=0.45\textwidth]{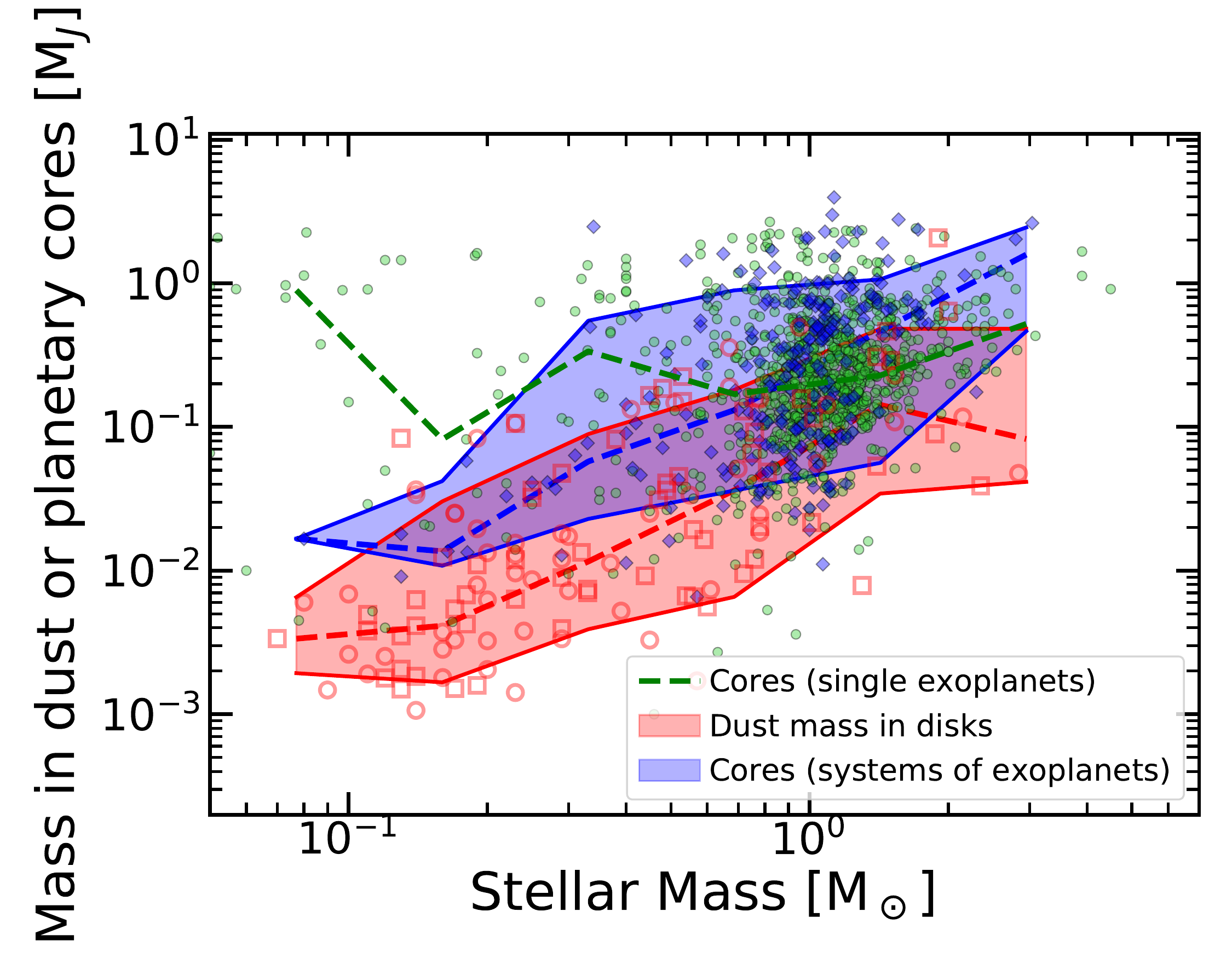}
\caption{Masses of the cores of single exoplanets and the sum of the cores in exoplanetary systems, as well as disk masses, as a function of the mass of their host star. The colored regions encompass the 10th and 90th percentiles of the distributions, while the dashed lines represent the median of the distributions, as labeled.
     \label{fig::mcore_mdust_mstar_perc}}
\end{figure}

We therefore perform a comparison of the disk dust masses with the core masses in planets and planetary systems. The main reasons for this choice is that the mass of planetary cores is mainly composed of the same heavy elements as the dusty component of disks, and that the disk dust masses are in general less uncertain than total disk masses. If the dust material in the disk has not yet grown to sizes larger than those probed by the millimeter flux of the disks, the disk dust material represents the material available to form planetary cores and rocky planets. The opposite possibility is considered in Sect.~\ref{sect::cores}.
For rocky planets, assumed here to be those with $M_{\rm pl} < 10 M_\oplus$, we use their total mass as a tracer of the heavy element content. For more massive planets, instead, we convert the planet mass into core masses using the relation found by \citet{thorngren16} using a sample of 47 transiting planets. We calculate the total mass of  heavy elements  in exosystems by summing up the individual masses of rocky planets and cores of giant planets in each exosystem. The dependence of the total core planets masses on stellar mass is shown in Fig.~\ref{fig::mcore_mdust_mstar_perc} together with the disk dust masses. 
The median values of the disk dust masses are systematically $\sim$0.2--0.3 times the heavy element content of planetary systems at all stellar masses $<$2 $M_\odot$. This confirms that the result obtained for disks and planets around low-mass stars is valid at all masses, and that the measured dust content in protoplanetary disks is smaller than the masses of the cores of exosystems.
It is also worth noting that planetary systems are concentrated within a few au from the central star, whereas most of the disk mass is farther out. This makes the imbalance between planet masses and disk masses even more pronounced, as planet masses are much larger than the local disk masses. Planet migration and/or pebble accretion might help to alleviate this issue because both processes concentrate the solid mass from a large portion of the disk into its inner part. In the next section we address possible solutions to this conundrum.

\section{Discussion}

The results shown in Sect.~\ref{sect::results} pose a serious question of whether the protoplanetary disks have enough mass to form planets when they are $\sim$1--3 Myr old, the age of the disks analyzed here. From the theoretical perspective, all models to explain the formation of planetesimals and planets are based on processes that are quite inefficient. 
In scenarios like pebble accretion \citep[e.g.,][]{johansen07,JL17,ormel17}, for example, only a small fraction of the pebble flux is captured by growing planets \citep[see][]{guillot14}. 
If the total disk dust mass, calculated assuming a distribution of grain sizes, is a proxy of the amount of both small dust particles and pebbles,  the disk dust mass is expected to be larger by at least one order of magnitude than the final mass in heavy elements in the planetary system. As we showed in Fig.~\ref{fig::mcore_mdust_mstar_perc}, the heavy element  content in exoplanetary systems can be as high as $\sim$2--3 $M_J$ for solar mass stars, and 0.02 $M_J$, about 6.4 $M_\oplus$, around brown dwarfs \citep[e.g., the TRAPPIST-1 system,][]{trappist}. In the solar system, summing up the mass in the terrestrial planets, in the cores of the giant planets, and in the progenitors of the Kuiper belt and Oort cloud, the total amount of heavy elements is about 130 $M_\oplus$, or 0.4 $M_J$ \citep[e.g.,][]{guillot14}. On the contrary, the highest measured dust masses of protoplanetary disks at $\sim$1--3 Myr are $\sim$0.5--1 $M_J$ around solar mass stars, and 0.01 $M_J$, $\sim$3 $M_\oplus$, around brown dwarfs (see also discussion by \citealt{pascucci16}). These are also the masses of a tiny fraction of disks, whereas the bulk of the disk population has masses that are one order of magnitude smaller (see Fig.~\ref{fig::mcore_mdust_mstar_perc}). 
This comparison shows that either planet accretion is more efficient than current models suggest or that 
other scenarios are needed.

A first naive possibility to explain this discrepancy is that the current surveys of exoplanets are biased towards high-mass planets, and the bulk of the population of exoplanets is instead much less massive than the one we observed. However, disk surveys are only targeting the disks that are still massive at $\sim$1--3 Myr, and, for example, they do not include less massive disks which are optically thin in the near-infrared (Class~III). These two biases should balance out.

Another possibility to explain the discrepancy is that disk dust masses are highly underestimated. It has been shown that  some dust mass can be confined into optically thick inner regions of disks  \citep{tripathi17,ansdell18} and that the opacity of disks and their temperature is still debatable \citep[e.g.,][]{andrews13,pascucci16}. 
It is also possible that a significant fraction of the dust content of disks has grown to cm sizes or more  \citep[e.g.,][]{williams12,NK14}, and is thus not detected by millimeter observations. 
However, there are reasons to believe that disk masses are not underestimated by more than one order of magnitude. For example, significantly higher disk masses would make the measurements of faint CO emission lines \citep[e.g.,][]{miotello17,long17} implying a gas-to-dust ratio much smaller than 100 even more difficult to explain, except in very old disks transitioning towards the debris phase.
Furthermore, significantly underestimated disk dust masses would conflict with the observed general agreement of the measured disk masses and mass accretion rates with expectations from viscous evolution theory \citep[e.g.,][]{manara16b,rosotti17,lodato17}. 
As discussed by \citet{pascucci16}, among others,  this possibility is  not enough to explain the apparent discrepancy between disk masses and planetary masses.

We do not believe that an efficient recycling of dust material in the disk is sufficient to explain the discrepancy. 
In such a scenario, the dust material that drifts inwards is then captured in disk winds originating in the innermost region of the disk ($R<1$ au) and is redeposited in the outer disk ($R>10$ au), where it will be able to again drift inwards. This scenario is commonly used to explain the observed population of calcium, aluminum-rich inclusions (CAI) and other aggregates present in bodies located throughout our solar system \citep[e.g.,][]{ciesla10,desch18}. Within this scenario, the single dust particle has multiple possibilities to be accreted onto planetesimals or planets.  
This increases the overall efficiency of planet accretion (i.e., the probability that dust is ultimately incorporated into a planet). However, even in the best-case scenario where the efficiency grows to $\sim$1, the total mass available for planet accretion would still be limited by $M_{\rm disk,dust}$, which as we have seen is typically smaller than the observed total planet heavy element mass. Moreover, it is implausible that all the material in the disk is recycled. Indeed, the ratio of mass loss in winds to mass accreted onto the central star sets a generous upper limit to the fraction of particles that can be recycled. Since this ratio is usually measured to be $<$50\%, and usually $\sim$10\% \citep[e.g.,][]{nisini18}, we consider that this scenario cannot explain the observed discrepancy between disk and planetary systems masses.

We propose in the following two possible explanations for the observations.

\subsection{Early formation of planetary cores}\label{sect::cores}
One way to explain the observations is to postulate that the cores of planets are formed in the very first Myr of the protoplanetary disk evolution, or even in the embedded phase while the disk is still forming. The disks for which masses have been measured have ages $>$1 Myr, and a general trend of declining disk mass with ages older than 1 Myr has been observed \citep[e.g.,][]{barenfeld16}. Thus, it is possible that disks were massive enough to form the cores of planets at younger ages.
This idea has been suggested by \citet{GR10}, \citet{williams12}, and \citet{NK14}, among others. In this scenario, the vast majority of the material composing planets must  already be in the form of planetesimals, for rocky planet formation, and of already-formed planetary cores. The latter condition is necessary as gas giant planets need to accrete gas from the gas-rich disk. Assuming a gas-to-dust ratio of 100, disks at $\sim$1--3 Myr have just the right amount of gas mass to explain the population of gas giants (see Fig.~\ref{fig::mplan_mdust_mstar}). Thus, cores must  already be in place at this age.
Scenarios have been proposed to explain that pebble accretion can form planetesimals very early ($<$0.1 Myr) in disks, when these are massive and possibly gravitationally unstable \citep{BC16}. 
However, it is expected that the formation of planetary cores is highly inefficient, with $\sim$350 $M_\oplus$ of pebbles needed to grow the core of Jupiter from half a lunar mass to 20 $M_\oplus$ \citep[e.g.,][]{morby16}. Thus, such an inefficiency would imply that disks were initially $\sim$10--100 times more massive than  is observed at ages $>$1 Myr. This would also imply that the vast majority of disks were initially gravitationally unstable. 
One issue of this scenario is that, if early disks are $\sim$10--100 times more massive than observed for disks older than 1 Myr, an extremely efficient gas-removal mechanism has to be found, consistent with the observations of Class~0 outflows \citep{frank14}.

This scenario is nevertheless in line with the recent result that the proto-core of Jupiter, with a mass of about 20$M_\oplus$, formed very rapidly, within 1 Myr \citep{kruijer17}. 
Also, the rings observed in the disk around the still embedded $<$1 Myr old HL Tau protostar \citep{hltau} may be carved by the presence of planets \citep[e.g.,][]{dipierro15} suggesting, again, early planet formation. It should  be noted that  processes other than planet formation have also been invoked to explain the formation of these ringed structures in HL Tau \citep[e.g.,][]{2015MNRAS.453L..78L}.

\subsection{Disks as conveyor belts}

We explore here another possibility where the disks are similar to a conveyor belt that transports material from the environment to the central star. In this scenario, the disk is replenished with material from the surrounding interstellar medium either continuously or in various episodes \citep[e.g.,][]{TB08,kuffmeier17}. This material, consisting of gas and small dust grains, is processed in the disk while drifting towards the central star. The material that is not processed in the disk is either accreted onto the central star or ejected through winds, and can even be partially recycled. 
This scenario is very similar to the one explored by \citet{padoan14}  to explain the luminosity problem of embedded protostars, the dependence of the mass accretion rate onto the central star, and the presence of large grains in disks. Starting from the modeling of stellar cluster formation in turbulent molecular clouds, \citet{padoan14} have shown that the inferred infall rates onto the star-disk system, a sink particle in their simulation, are comparable to the mass accretion rates on the central star measured spectroscopically in young stellar objects \citep[$\sim 10^{-12} - 10^{-6} M_\odot$/yr,][]{manara12}, covering almost this whole range of values at any \mstar. A similarly vigorous accretion rate ($\sim$4$\times 10^{-8} M_\odot$/yr) of material from the environment, in this case an envelope, onto the protoplanetary disk was measured by \citet{semenov05} for the AB Aur young stellar object, which has \mstar$\sim$2.4$M_\odot$. 
Assuming an accretion rate of $10^{-8} M_\odot$/yr, which is typical for all \mstar\ in the simulations if such a process of accretion of material onto the disk from the environment is sustained for $\sim$ 1 Myr after the time we are observing them, this would give to the disk 0.1 $M_J$ of additional dust material (and 10 $M_J$ of gas) to be used for planet formation. This value is comparable to the median of the measured disk dust masses for disks around stars with $M_\star \gtrsim 1 M_\odot$, similar to the AB Aur system. Disks around lower mass stars will have in general a lower accretion rate onto the disk, since in the Bondy-Hoyle accretion scenario \macc$\propto$\mstar$^2$, but the scatter of the relation between the accretion rate from the environment onto the disk as a function of \mstar \ has a large spread \citep{padoan14}, and it is thus possible that this process gives enough material to the disks to form planets. 
Simulations of face-on accretion of ISM material onto protoplanetary disks indeed show that the rate of material accreted onto the disk can  even be as high as 10$^{-6} M_\odot$/yr depending on the density of the ISM material, on the relative velocity between the disk and the ISM, and on the disk sizes \citep{wijnen16,wijnen17}.
However, it is still unclear how many star-disk systems are actually accreting material from the environment and whether this process is continuous or episodic.

Finally, it is worth noting that \citet{scicluna14} have even proved that disk-less stars going through over-densities of the local molecular cloud remnant are able to capture enough material to form a new disk around them, thus giving planet formation a second chance.

\section{Conclusions}

We  collected the exoplanet and exoplanetary system masses and the masses of their host stars, and recalculated protoplanetary disk and disk-host star masses using the newly estimated distances based on Gaia parallaxes. We observe that single exoplanets around very low-mass stars can have masses almost as high as their host stars, whereas this is never observed in exoplanetary systems. This possibly points towards a different formation mechanism for these massive exoplanets around very low-mass stars. 
We have shown that current measurements of protoplanetary disk dust masses in $\sim$1--3 Myr old regions are lower than or, at most, comparable to the amount of heavy element material in the exoplanetary systems discovered to date. Unless disk dust masses are underestimated by more than one order of magnitude, this result challenges the current theories of planet formation according to which planet formation should not be a very efficient process. We discuss two scenarios to explain this discrepancy: 
(i) an early formation of planetary cores  at ages $<0.1-1$ Myr, when the disks may be more massive than those, much older, that are typically observed and (ii) the possibility that disks are replenished by fresh material from the environment during their lifetimes so that the total amount of material available for planet formation greatly exceeds  that observed in the disk at any given time.

In order to confirm either of these hypotheses it is important to obtain measurements of disk dust masses and measurements of dust size distributions in disks in younger regions and in more embedded objects (Class 0/I), and to determine how many disks are replenished with material from the environment throughout their lifetimes.

\begin{acknowledgements}
We thank the referee, Jonathan Williams, for the constructive and useful report that helped  improve the manuscript.
CFM acknowledges support from the Laboratoire Lagrange and the Observatoire de la C\^ote d'Azur for a scientific visit during which this work was started. CFM acknowledges support through the ESO fellowship. This work made use of the Python packages Numpy and matplotlib. We acknowledge inspiring discussions with P. Hennebelle and G. Rosotti. This work has made use of data from the European Space Agency (ESA) mission {\it Gaia} (\url{https://www.cosmos.esa.int/gaia}), processed by the {\it Gaia}
Data Processing and Analysis Consortium (DPAC, \url{https://www.cosmos.esa.int/web/gaia/dpac/consortium}). Funding for the DPAC has been provided by national institutions, in particular the institutions
participating in the {\it Gaia} Multilateral Agreement.
\end{acknowledgements}

%
%

\appendix

\section{Revised estimates of disk and stellar masses}\label{sect::dr2}

In the light of the recent Data Release 2 \citep{gaiadr2} of the Gaia mission \citep{gaia}, we have revised the stellar masses and disk masses for all the targets in the samples of young stars with disks in the Lupus and Chamaeleon~I star-forming regions. The data for the objects in these regions were initially analyzed assuming a distance of 150 pc or 200 pc for the targets in the Lupus region \citep{alcala17,ansdell16} and 160 pc for targets in the Chamaeleon~I region \citep{manara17a,pascucci16}. Here we queried the Gaia archive for all these targets to obtain the measured parallaxes and their uncertainties. We  inverted the parallax measured in milliarcseconds to obtain the distance of the targets in parsec in the cases when the relative uncertainty on the parallax measurement was smaller than 10\%. For the other targets with either more uncertain measurements of parallaxes or without a Gaia parallax measurement, we have assumed as a distance the inverted weighted mean of the parallaxes of all the objects in either the Lupus or Chamaleon~I samples. These median distances are 158.5 pc and 190 pc for the two regions, respectively. We note that the latter is well in line withe the estimate of \citet{voirin18} based on Gaia DR1 data. The distances used here are reported in Col. 3 of Tables~\ref{tab::samplelup} and \ref{tab::samplecha}. 

The new estimates of stellar masses are obtained by rescaling the stellar luminosities from \citet{alcala17} and \citet{manara17a} to the new distances, and by comparing the new stellar luminosities and the effective temperatures of the stars with the evolutionary models by \citet{B15}, when possible, or \citet{S00} for stellar masses higher than 1.4 $M_\odot$. These newly determined stellar masses are reported in Co. 4 of Tables~\ref{tab::samplelup} and \ref{tab::samplecha}. Finally, the disk masses are obtained by rescaling the disk masses reported by \citet{ansdell16} and \citet{pascucci16} for the new distances, and are reported in Col. 5 of Tables~\ref{tab::samplelup} and \ref{tab::samplecha}. 

\begin{figure}[]
\centering
\includegraphics[width=0.45\textwidth]{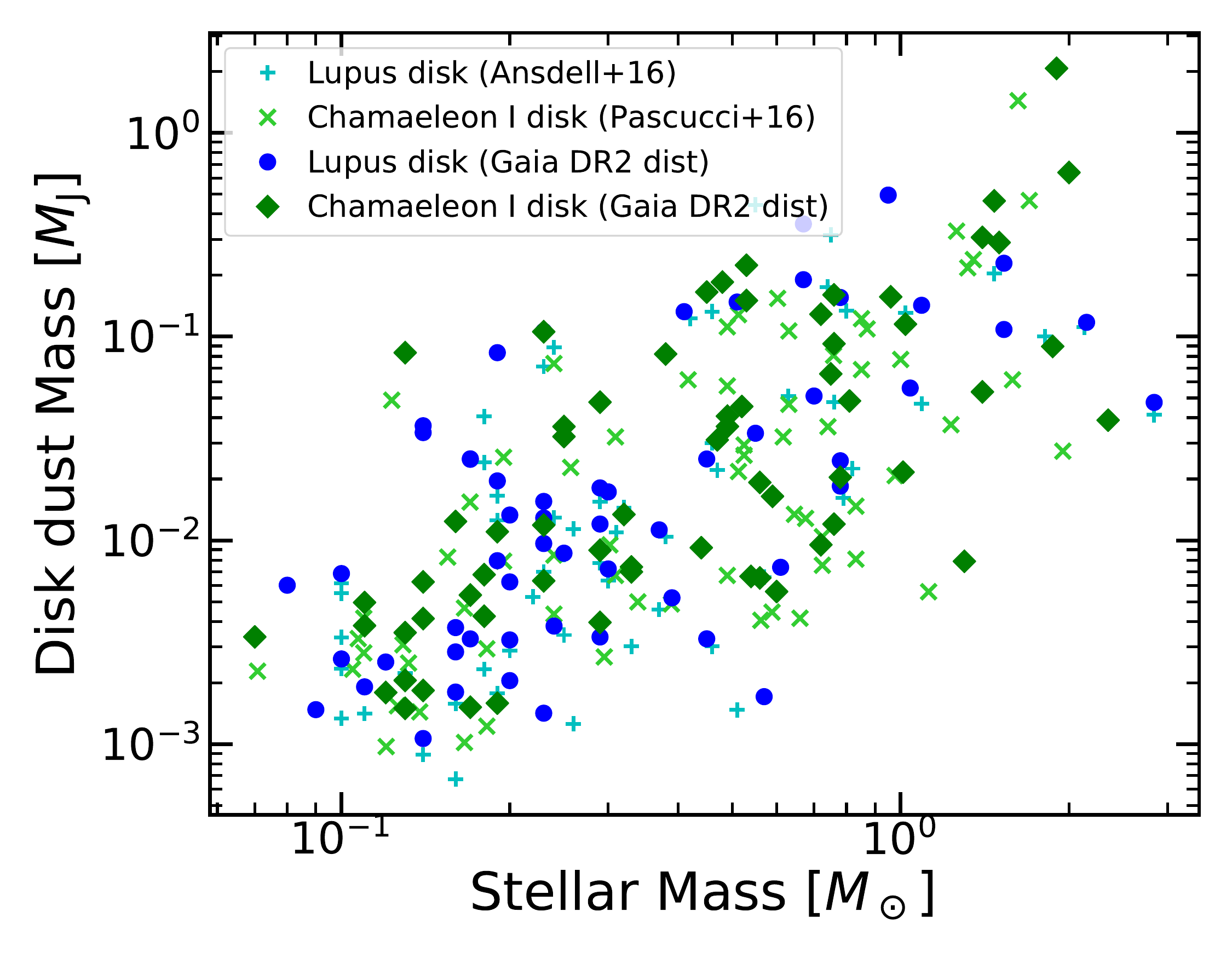}
\caption{Disk mass vs stellar mass with new distances.
     \label{fig::mdisk_mstar_dr2}}
\end{figure}

As shown in Fig.~\ref{fig::mdisk_mstar_dr2}, the new distances do not drastically modify the overall distribution of the values of disk masses as a function of stellar masses with respect to previous works. Overall, we note that disk masses in the Chamaeleon~I region are slightly higher than previously reported, while a slightly smaller spread is observed in the disk mass distribution for the disks in the Lupus region.

\begin{table*}  
\begin{center} 
\footnotesize 
\caption{\label{tab::samplelup} Stellar and disk masses for the objects in the Lupus star-forming region } 
\begin{tabular}{l|l| ccc   } 
\hline \hline 
Name & Other  &  dist & $M_\star$ & $M_{\rm disk,dust}$    \\  
 & Names  &  [pc] & [$M_\odot$] & [$M_\oplus$]    \\  
\hline 
Sz65                     & Sz65                     & 155.29   &  0.70     &  16.2437  $\pm$  0.0806     \\
Sz66                     & Sz66                     & 157.34   &  0.29    &   3.8218   $\pm$  0.0750     \\
J15450887-3417333        & SSTc2dJ154508.9-341734   & 154.96   &  0.14     &  11.6050  $\pm$  0.1254     \\
Sz68                     & Sz68                     & 154.19   &  2.15     &  37.3405  $\pm$  0.1142     \\
Sz69                     & Sz69                     & 154.55   &  0.20    &   4.2313   $\pm$  0.0699     \\
Sz71                     & Sz71                     & 155.89   &  0.41     &  42.1459  $\pm$  0.1600     \\
Sz72                     & Sz72                     & 155.89   &  0.37    &   3.5790   $\pm$  0.0711     \\
Sz73                     & Sz73                     & 156.78   &  0.78    &   7.8125   $\pm$  0.1413     \\
Sz74                     & Sz74                     & 158.50   &  0.30    &   5.4946   $\pm$  0.0709     \\
Sz81A                    & Sz81                     & 159.86   &  0.19    &   2.5251   $\pm$  0.0480     \\
Sz82                     & Sz82                     & 158.45   &  0.95      & 157.3413  $\pm$ 23.6012    \\
Sz83                     & Sz83                     & 159.57   &  0.67      & 113.5365 $\pm$  0.1915     \\
Sz84                     & Sz84                     & 152.64   &  0.17    &   7.9432   $\pm$  0.0973     \\
Sz129                    & Sz129                    & 161.68   &  0.78     &  49.4522  $\pm$  0.1420     \\
RYLup                    & RYLup                    & 159.10   &  1.53     &  72.8398  $\pm$  0.3067     \\
J16000060-4221567        & SSTc2dJ160000.6-422158   & 161.20   &  0.20    &   0.6514   $\pm$  0.0516     \\
J16000236-4222145        & SSTc2dJ160002.4-422216   & 164.17   &  0.23     &  33.7391  $\pm$  0.1774     \\
J16002612-4153553        & SSTc2dJ160026.1-415356   & 164.33   &  0.14    &   0.3385   $\pm$  0.0536     \\
Sz130                    & Sz130                    & 160.27   &  0.39    &   1.6607   $\pm$  0.0940     \\
MYLup                    & MYLup                    & 156.59   &  1.09     &  45.2837  $\pm$  0.1946     \\
Sz131                    & Sz131                    & 160.31   &  0.30    &   2.3005   $\pm$  0.0779     \\
Sz133                    & Sz133                    & 153.08   &  \nodata   &  16.9009  $\pm$  0.1885     \\
Sz88A                    & Sz88                     & 158.43   &  0.61    &   2.3438   $\pm$  0.0786     \\
J16070384-3911113        & SSTc2dJ160703.9-391112   & 158.50   &  0.16    &   1.1860   $\pm$  0.1443     \\
Sz90                     & Sz90                     & 160.40   &  0.78    &   5.8663   $\pm$  0.1236     \\
J16073773-3921388        & Lup713                   & 174.40   &  0.11    &   0.6068   $\pm$  0.0572     \\
Sz95                     & Sz95                     & 158.17   &  0.29    &   1.0661   $\pm$  0.0470     \\
J16080017-3902595        & Lup604s                  & 159.89   &  0.12    &   0.8038   $\pm$  0.0454     \\
Sz96                     & Sz96                     & 156.55   &  0.45    &   1.0444   $\pm$  0.0717     \\
J16081497-3857145        & 2MASSJ16081497-3857145   & 158.50   &  0.10    &   2.1832   $\pm$  0.0787     \\
Sz97                     & Sz97                     & 157.75   &  0.24    &   1.2061   $\pm$  0.0468     \\
Sz98                     & Sz98                     & 156.22   &  0.67     &  60.4867  $\pm$  0.3620     \\
Sz100                    & Sz100                    & 136.94   &  0.14     &  10.7435  $\pm$  0.1136     \\
Sz103                    & Sz103                    & 159.50   &  0.23    &   3.0664   $\pm$  0.0744     \\
J16083070-3828268        & SSTc2dJ160830.7-382827   & 156.12   &  1.53     &  34.4472  $\pm$  0.2851     \\
Sz104                    & Sz104                    & 165.46   &  0.16    &   0.9008   $\pm$  0.0543     \\
V856Sco                  & V856Sco                  & 161.00   &  2.84     &  15.1102  $\pm$  0.0758     \\
Sz106                    & Sz106                    & 161.67   &  0.57    &   0.5433   $\pm$  0.0546     \\
Sz108B                   & Sz108B                   & 168.99   &  0.17    &   7.9851   $\pm$  0.1015     \\
J16084940-3905393        & Par-Lup3-3               & 159.30   &  0.23    &   0.4506   $\pm$  0.0450     \\
V1192Sco                 & Par-Lup3-4               & 150.81   &  \nodata  &   0.2162   $\pm$  0.0451     \\
Sz110                    & Sz110                    & 159.51   &  0.23    &   4.0926   $\pm$  0.0771     \\
J16085324-3914401        & 2MASSJ16085324-3914401   & 167.71   &  0.29    &   5.7493   $\pm$  0.0823     \\
J16085373-3914367        & 2MASSJ16085373-3914367   & 158.50   &  0.10    &   0.8318   $\pm$  0.0499     \\
Sz111                    & Sz111                    & 158.33   &  0.51     &  46.9712  $\pm$  0.2566     \\
J16085529-3848481        & 2MASSJ16085529-3848481   & 157.52   &  0.09    &   0.4691   $\pm$  0.0466     \\
Sz112                    & Sz112                    & 160.27   &  0.17    &   1.0436   $\pm$  0.0456     \\
Sz113                    & Sz113                    & 163.24   &  0.19    &   6.2207   $\pm$  0.0751     \\
J16090141-3925119        & SSTc2d160901.4-392512    & 164.31   &  0.23    &   4.9349   $\pm$  0.1918     \\
Sz114                    & Sz114                    & 162.25   &  0.19     &  26.5093  $\pm$  0.1127     \\
J16092697-3836269        & SSTc2dJ160927.0-383628   & 159.35   &  0.20    &   1.0317   $\pm$  0.0716     \\
Sz117                    & Sz117                    & 158.56   &  0.25    &   2.7494   $\pm$  0.0473     \\
Sz118                    & Sz118                    & 163.90   &  1.04     &  17.7723  $\pm$  0.2722     \\
J16095628-3859518        & Lup818s                  & 156.93   &  0.08    &   1.9138   $\pm$  0.0463     \\
J16102955-3922144        & SSTc2dJ161029.6-392215   & 163.23   &  0.20    &   1.9870   $\pm$  0.0974     \\
Sz123A                   & Sz123                    & 158.50   &  0.55     &  10.6640  $\pm$  0.1470     \\
J16124373-3815031        & SSTc2dJ161243.8-381503   & 159.81   &  0.45    &   7.9707   $\pm$  0.1307     \\
J16134410-3736462        & SSTc2dJ161344.1-373646   & 160.00   &  0.16    &   0.5722   $\pm$  0.0802     \\
\hline 
\end{tabular} 
\tablefoot{Original stellar masses are from \citet{alcala14} and \citet{alcala17}. Original disk masses are from \citet{ansdell16}. The distances used to modify these values are  from \citet{gaiadr2}. } 
\end{center} 
\end{table*}

\begin{table*}  
\begin{center} 
\footnotesize 
\caption{\label{tab::samplecha} Stellar and disk masses for the objects in the Chamaeleon~I star-forming region } 
\begin{tabular}{l|l| ccc   } 
\hline \hline 
2MASS & Other  &  dist & $M_\star$ & $M_{\rm disk,dust}$    \\  
Name  & Names  &  [pc] & [$M_\odot$] & [$M_\oplus$]    \\  
\hline 
2MASSJ10533978-7712338   & ...                    &   191.81   &  0.33   &    2.2259  $\pm$   1.7065 \\     
2MASSJ10555973-7724399   & T3A                    &   185.08   &  0.81   &    15.3631  $\pm$  1.3908 \\     
2MASSJ10561638-7630530   & ESOHalpha553           &   196.48   &  0.14   &    1.9879  $\pm$   1.5696 \\     
2MASSJ10563044-7711393   & T4                     &   183.09   &  0.76   &    50.9436  $\pm$  1.3219 \\     
2MASSJ10574219-7659356   & T5                     &   190.00   &  0.32   &    4.2586  $\pm$   1.5444 \\     
2MASSJ10580597-7711501   & ...                    &   186.57   &  0.11   &    1.2118  $\pm$   1.4433 \\     
2MASSJ10581677-7717170   & T6                     &   189.84   &  1.47   &    147.4129  $\pm$ 1.4123 \\     
2MASSJ10590108-7722407   & T7                     &   185.21   &  0.76   &    29.3149  $\pm$  1.3753 \\     
2MASSJ10590699-7701404   & T8                     &   187.48   &  2.00   &    203.0813  $\pm$ 1.3752 \\     
2MASSJ11004022-7619280   & T10                    &   191.54   &  0.23   &    33.5953  $\pm$  1.4367 \\     
2MASSJ11022491-7733357   & T11                    &   176.26   &  1.50   &    92.0593  $\pm$  1.2175 \\     
2MASSJ11025504-7721508   & T12                    &   182.24   &  0.19   &    0.5047  $\pm$   1.4892 \\     
2MASSJ11040425-7639328   & CHSM1715               &   192.31   &  0.18   &    1.3482  $\pm$   1.5306 \\     
2MASSJ11040909-7627193   & T14                    &   191.78   &  0.96   &    49.8157  $\pm$  1.4450 \\     
2MASSJ11044258-7741571   & ISO52                  &   193.15   &  0.23   &    2.0116  $\pm$   1.5144 \\     
2MASSJ11045701-7715569   & T16                    &   194.46   &  0.29   &    1.2572  $\pm$   2.0343 \\     
2MASSJ11062554-7633418   & ESOHalpha559           &   209.30   &  0.13   &    26.5032  $\pm$  1.7167 \\     
2MASSJ11064180-7635489   & Hn5                    &   195.28   &  0.17   &    0.4820  $\pm$   1.7566 \\     
2MASSJ11065906-7718535   & T23                    &   190.34   &  0.25   &    11.5032  $\pm$  1.4359 \\     
2MASSJ11065939-7530559   & ...                    &   196.24   &  0.11   &    1.5752  $\pm$   1.5836 \\     
2MASSJ11071206-7632232   & T24                    &   195.75   &  0.54   &    2.1143  $\pm$   1.9855 \\     
2MASSJ11071860-7732516   & ChaHalpha9             &   198.58   &  0.13   &    0.4760  $\pm$   1.8295 \\     
2MASSJ11072074-7738073   & T26                    &   190.62   &  2.35   &    12.3622  $\pm$  1.5004 \\     
2MASSJ11074245-7733593   & ChaHalpha2             &   190.00   &  0.13   &    1.1201  $\pm$   1.6764 \\     
2MASSJ11074366-7739411   & T28                    &   194.81   &  0.45   &    52.5996  $\pm$  1.4903 \\     
2MASSJ11074656-7615174   & CHSM10862              &   194.22   &  0.07   &    1.0675  $\pm$   1.5858 \\     
2MASSJ11075730-7717262   & CHXR30B                &   187.10   &  0.44   &    2.9235  $\pm$   1.5474 \\     
2MASSJ11075792-7738449   & T29                    &   163.19   &  1.01   &    6.8730  $\pm$   1.1208 \\     
2MASSJ11075809-7742413   & T30                    &   184.46   &  0.29   &    2.8416  $\pm$   1.5023 \\     
2MASSJ11080148-7742288   & T31                    &   190.00   &  0.75   &    20.8577  $\pm$  1.4490 \\     
2MASSJ11080297-7738425   & ISO126                 &   190.00   &  0.53   &    47.7823  $\pm$  1.4183 \\     
2MASSJ11081509-7733531   & T33A                   &   190.00   &  1.40   &    97.5590  $\pm$  1.4131 \\     
2MASSJ11083905-7716042   & T35                    &   188.36   &  0.78   &    6.4824  $\pm$   1.5029 \\     
2MASSJ11085367-7521359   & ...                    &   188.27   &  0.49   &    11.5165  $\pm$  1.4639 \\     
2MASSJ11085464-7702129   & T38                    &   186.01   &  0.60   &    1.7817  $\pm$   1.6551 \\     
2MASSJ11092266-7634320   & C1-6                   &   203.25   &  0.56   &    2.0788  $\pm$   1.9762 \\     
2MASSJ11092379-7623207   & T40                    &   192.31   &  0.48   &    58.8523  $\pm$  1.4513 \\     
2MASSJ11094621-7634463   & Hn10E                  &   195.04   &  0.33   &    2.3551  $\pm$   1.7559 \\     
2MASSJ11094742-7726290   & B43                    &   192.96   &  0.53   &    71.2352  $\pm$  1.4628 \\     
2MASSJ11095340-7634255   & T42                    &   201.96   &  0.72   &    40.9535  $\pm$  1.6319 \\     
2MASSJ11095407-7629253   & T43                    &   190.00   &  0.52   &    14.4300  $\pm$  1.4688 \\     
2MASSJ11095873-7737088   & T45                    &   191.29   &  0.47   &    9.8888  $\pm$   1.4691 \\     
2MASSJ11100010-7634578   & T44                    &   192.08   &  1.90   &    658.7556  $\pm$ 1.4422 \\     
2MASSJ11100369-7633291   & Hn11                   &   201.01   &  0.59   &    5.2263  $\pm$   1.7104 \\     
2MASSJ11100469-7635452   & T45A                   &   195.01   &  0.76   &    3.8183  $\pm$   1.6431 \\     
2MASSJ11100704-7629376   & T46                    &   179.61   &  0.72   &    3.0229  $\pm$   1.4048 \\     
2MASSJ11101141-7635292   & ISO237                 &   195.37   &  1.02   &    36.5996  $\pm$  1.5194 \\     
2MASSJ11103801-7732399   & CHXR47                 &   190.00   &  1.30   &    2.5077  $\pm$   1.6307 \\     
2MASSJ11104959-7717517   & T47                    &   185.16   &  0.38   &    26.1129  $\pm$  1.3729 \\     
2MASSJ11105333-7634319   & T48                    &   194.68   &  0.29   &    15.1497  $\pm$  1.5435 \\     
2MASSJ11105359-7725004   & ISO256                 &   195.77   &  0.16   &    3.9378  $\pm$   1.5630 \\     
2MASSJ11105597-7645325   & Hn13                   &   140.23   &  0.12   &    0.5694  $\pm$   0.8777 \\     
2MASSJ11111083-7641574   & ESOHalpha569           &   190.00   &  \nodata   &  25.6606  $\pm$  1.4564 \\     
2MASSJ11113965-7620152   & T49                    &   190.51   &  0.25   &    10.2706  $\pm$  1.4716 \\     
2MASSJ11114632-7620092   & CHX18N                 &   192.51   &  1.40   &    17.0085  $\pm$  1.5003 \\     
2MASSJ11120351-7726009   & ISO282                 &   185.49   &  0.14   &    1.3134  $\pm$   1.4191 \\     
2MASSJ11120984-7634366   & T50                    &   193.23   &  0.18   &    2.1573  $\pm$   1.5650 \\     
2MASSJ11122772-7644223   & T52                    &   193.24   &  1.87   &    28.4416  $\pm$  1.4909 \\     
2MASSJ11123092-7644241   & T53                    &   196.00   &  0.56   &    6.1133  $\pm$   1.6065 \\     
2MASSJ11132446-7629227   & Hn18                   &   189.52   &  0.23   &    3.7763  $\pm$   1.5473 \\     
2MASSJ11142454-7733062   & Hn21W                  &   188.95   &  0.19   &    3.5031  $\pm$   1.4600 \\     
2MASSJ11160287-7624533   & ESOHalpha574           &   190.00   &  \nodata   &  6.0154  $\pm$   1.6076 \\     
2MASSJ11173700-7704381   & T56                    &   188.38   &  0.49   &    12.9369  $\pm$  1.4509 \\     
2MASSJ11183572-7935548   & ...                    &   94.62   &   0.17   &    1.7129  $\pm$   0.3583 \\     
2MASSJ11241186-7630425   & ...                    &   184.75   &  0.13   &    0.6530  $\pm$   1.4867 \\     
2MASSJ11432669-7804454   & ...                    &   180.63   &  0.14   &    0.5826  $\pm$   1.8409 \\     
\hline 
\end{tabular} 
\tablefoot{Original stellar masses are from \citet{manara16a} and \citet{manara17a}. Original disk masses are from \citet{pascucci16}. The distances used to modify these values are  from \citet{gaiadr2}. } 
\end{center} 
\end{table*}  

\end{document}